\documentclass{PoS}
\newcommand{\ar}{\arrowvert}
\newcommand{\ra}{\rangle}

\usepackage{bm}

\title{How to identify gluonium states\\ using QCD counting rules}

\ShortTitle{Identifying gluonium states}

\author{\speaker{Felipe J. Llanes-Estrada}\thanks{We thank the organizers of Confinement XIII for an inspiring venue, warm welcome, and recognizing this work with the Phenomenology Poster Prize; and Richard Lebed and Jose R. Pel\'aez for helpful discussions. Supported by Spanish grant MINECO:FPA2016-75654-C2-1-P and US Dept. of Energy Contract DE-AC02-76SF00515. 
\ \ \ \ \ \ \ \ \ \ \ \ \ \ \ \ \ \ \ \ \ \ \ \ \ \ \ \ \ \ \ \ \ 
SLAC preprint  SLAC-PUB-17359.
}\\
        Depto. de F\'{\i}sica Te\'orica, Univ. Complutense de Madrid, 28040 Spain\\
        E-mail: \email{fllanes@fis.ucm.es}}
        
         \author{Stanley J. Brodsky \\
        SLAC National Accelerator Laboratory, Stanford University, Stanford, CA 94025, USA\\
        E-mail: \email{sjbth@slac.stanford.edu}}

\abstract{
Verifying the existence of bound states of gluons and distinguishing them from conventional quark--antiquark, hybrid or tetraquark states has remained a key problem in QCD.
We show that QCD counting rules for the power-law fall-off of production  cross sections at high momentum transfer can be used  to distinguish gluonium states from conventional hadrons.
The valence two-gluon contribution to  a
$0^+$ gluonium  bound state  has $L=0$ and thus twist (dimension minus spin of their minimum interpolating operators)  $\tau=2$.
The competing twist assignments for scalar  $f_0$ mesons  have  twist
$\tau = 3$ for the valence $|q \bar q \rangle $ configuration or $|q\bar{q}g\rangle$ in an s-wave, and  $\tau \ge 4$ for 
 $|q  q \bar q \bar q \rangle$ tetraquarks, etc.  
Thus, the production cross section for mesons with quark--containing valence wavefunctions
 relative to glueball production should be suppressed by at least a power of momentum transfer.   
Distinguishing these processes is feasible in 
exclusive $e^-e^+ \to \phi f_0$ reactions at 9 and 11 GeV center of mass energy at Belle-II. 
In the case of single--particle inclusive hadroproduction $ A B \to C X$, the cross section for scalar gluonium production at high transverse momentum $p_T$ and fixed $x_T = 2 {p_T\over \sqrt s} $ will dominate meson or tetraquark production  by at least two powers of $p_T$.  }

\FullConference{XIII Quark Confinement and the Hadron Spectrum - Confinement2018\\
		31 July - 6 August 2018\\
		Maynooth University, Ireland}

\begin{document}

Bound states of two or three color-octet gluons into color-singlet hadrons -- ``gluonium"  appears to be an essential prediction of the color-confining $SU(3)_C $ QCD Lagrangian. 
The simplest such ``glueball"  $|gg>$ bound state will have $J^{PC} = 0^{++}$ and zero isospin.
It has been 24 years since the first \emph{Confinement \& the Hadron Spectrum} conference; however
the existence or absence of a ``glueball'' has not yet been convincingly established  --  and not for want of searching. 
Various hadron experiments~\cite{Tanabashi:2018oca} have established five scalar $f_0$ mesons in the 1--2 GeV energy range  where  the glueball is generally expected~\cite{Brambilla:2014jmp}.  

In fact, not all approaches to QCD predict gluonium states.  For example, the superconformal algebra approach~\cite{Dosch:2015nwa} gives a comprehensive description of the hadron spectrum in terms of multiplets of mesons, baryons, and tetraquarks without gluonia. In this case the strongly-interacting QCD gluon degrees of freedom in the non-perturbative regime are subsumed into the color confinement potential.

The search for gluonium states is greatly complicated by the fact that a valence  $|gg>$ Fock state can mix with QCD Fock states containing quark pairs with the same quantum numbers. 
The contemporary problem is thus how to discriminate between quark-based versus gluon based mesons, and how to  avoid the model dependence of conventional spectroscopic mixing  to make the most of data.   One regime in which firm statements can be made is to analyze the dynamical dependence of production cross sections when all scales become large~\cite{Brodsky:2018snc}.    For related studies see refs.~\cite{Brodsky:2015wza,Brodsky:2016uln,Brodsky:2017icd}.

Exclusive scattering processes such as $AB\to CD$ at fixed CM angle and large energy, have
differential cross sections that scale~\cite{Brodsky:1973kr,Matveev:1973ra} as a power-law in $s$, 
\begin{equation} \label{differentialcounting}
\frac{d\sigma(AB\to CD)}{dt}  = {f(\theta_{CM})\over {s^{n-2}}} .
\end{equation}
Here, $n=n_i+n_f$ is the total minimum number of pointlike particles composing the initial and final state ones (that is, the fundamental components of $A\dots D$ if they are not fundamental themselves).
The well--known equation~(\ref{differentialcounting}) can be extended to  include internal orbital angular momentum~\cite{Amati:1968kr,Ciafaloni:1968ec,Brodsky:1974vy}. Generically, bound states see a short--distance suppression due to centrifugal factors which is reflected in the twist of the leading bound-state amplitude.
Nonrelativistic Schr\"odinger wavefunctions scale as $r^L$; relativistic Bethe-Salpeter wavefunctions contain also such suppression as do front--form wavefunctions in terms of the boost-invariant ``radial" variable $\zeta$. 

This suppression  carries over to amplitudes in which a hadron with $L$ units of internal angular momentum participates, so they fall as $\left({\sqrt s}\right)^{-L}$~\cite{Brodsky:1981kj}, and the cross sections by $s^{-L}$, becoming
\begin{equation} \label{counting}
\frac{d\sigma}{dt}  = {f(\theta_{CM})\over {s^{n +L -2}}}
\end{equation}
(here all the internal orbital angular momenta are summed in $L$).

Thus, for the reaction $e^-e^+\to \phi f_0$ one counts $n_i=2$ (two leptons in the initial state) and $n_f=4$ or more in the final state (the $\phi$ contributes one quark and one antiquark with $L=0$, and the $f_0$ a minimum of two units if a $\ar gg\ra$ glueball). Conventional quark--antiquark mesons as well as other exotic configurations are suppressed by the powers shown in table~\ref{tab:counting}.

\begin{table}[b]
\caption{Power of $s$ suppressing candidate wavefunctions \emph{relative to the glueball} in reactions at large momentum transfer involving an $f_0$ meson.\label{tab:counting}}
\begin{center}
\begin{tabular}{|c|cccc|} \hline
Wavefunction& $gg$  & $q\bar{q}\arrowvert_{L=1}$ & $q\bar{q}g$ & $q\bar{q} q\bar{q}$ \\
$n_f+L$     &  2    &  3         & 3         & 4 \\               
Suppression &  1    & $s^{-1}$   & $s^{-1}$  & $s^{-2}$ \\
\hline 
\end{tabular}
\end{center}
\end{table}

QCD predicts that the power--law in Eq.~(\ref{counting}) acquires logarithmic corrections~\cite{Lepage:1980fj,Efremov:1979qk}. These have been exposed in an explicit calculation of $\gamma\gamma\to \pi_0 f_2$ with $f_2$ treated as a tensor glueball~\cite{Kivel:2017tgt} and found to be manageably small.
\footnote{This reaction is a bit less convenient as the one that we propose because, first, one needs two photons, and second, the tensor glueball is expected to be heavier than the scalar one by about half a GeV~\cite{Szczepaniak:1995cw}, so that the conformal regime where $E\gg m$ is achieved only at a higher scale.}.

\begin{figure}
\includegraphics[width=0.4\columnwidth]{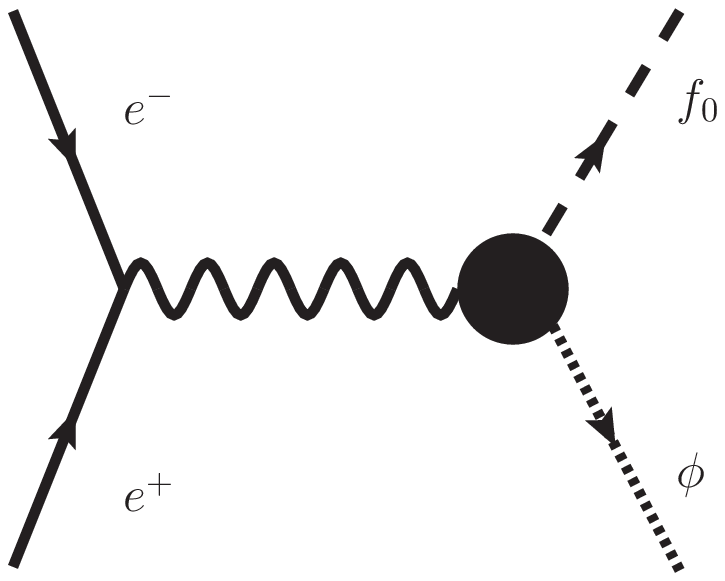}\hspace{1cm}
\includegraphics[width=0.45\columnwidth]{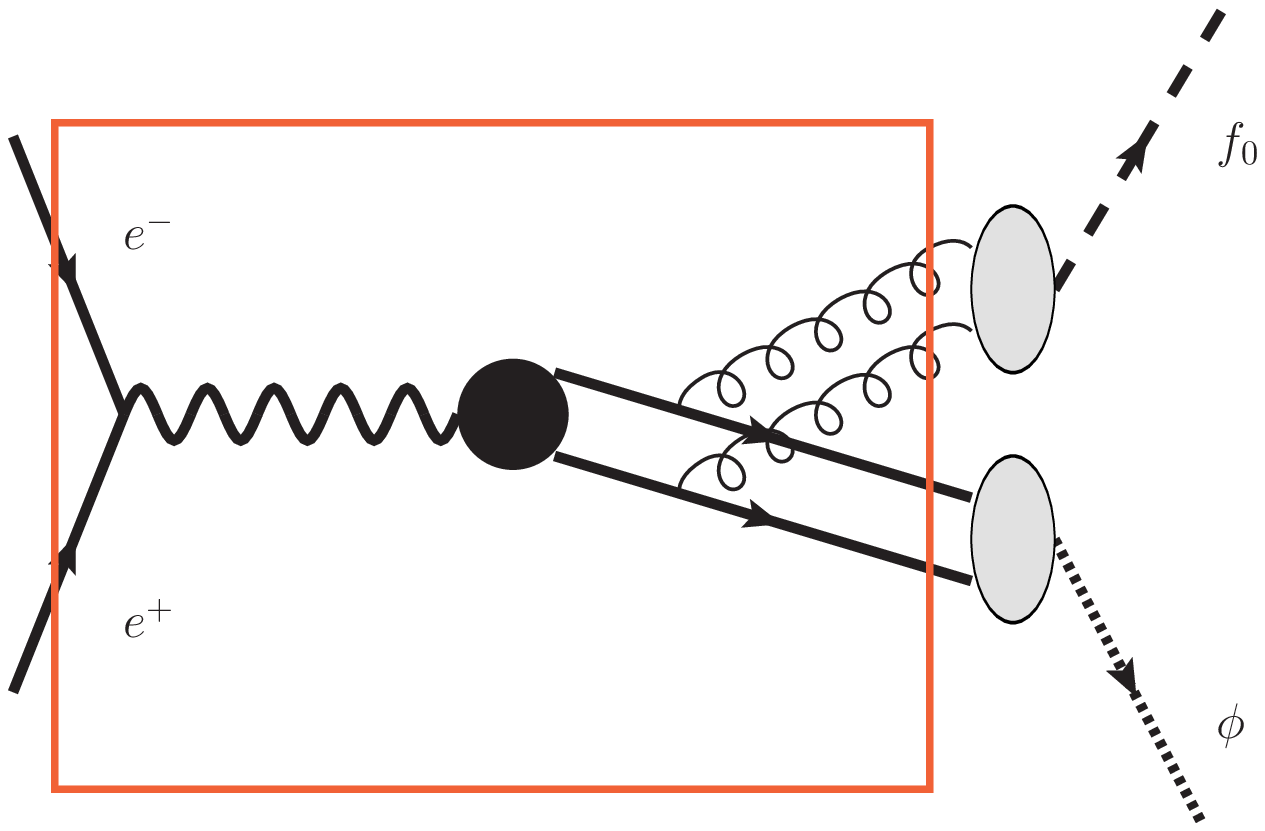}
\caption{\label{fig:Feynmanhadrons} 
The left plot depicts the process $e^-e^+\to\phi f_0$ at the hadron level, while the right plot
breaks this down to the quark--gluon level; the counting rules read off the number of underlying fundamental fields, as shown by the  box, red online, where it intersects the diagram.
The final state mesons eventually yield easily identifiable pairs of charged tracks $K^-K^+$ and $\pi^+\pi^-$. }
\end{figure}

We show the basic Feynman diagrams for the reaction $e^-e^+\to \phi f_0$ in figure~\ref{fig:Feynmanhadrons}. 

The counting rule of Eq.~(\ref{counting}) applied to the right plot of figure~\ref{fig:Feynmanhadrons} then returns  for any glueball produced among the $f_0$s a differential $\frac{d\sigma}{dt}  = f(\theta) \frac{1}{s^4}$. But summing all events in a fixed solid angle covering the barrel detector (where $t\sim s$ so all scales are large) lowers the power by one, resulting in
\begin{eqnarray}\label{glueballscaling}
\sigma \left(f_0=\ar \bf{gg} \ra +\dots \right) & \sim & \frac{\rm constant}{\bf s^3} \\ \nonumber
  & \phantom{\sim} & \\ \nonumber
\sigma \left(f_0=\ar {\bf{q\bar{q}}}\ra_{L=1} +\dots \right)       & \sim& \frac{\rm constant}{\bf s^4}   \\ \nonumber
  & \phantom{\sim} & \\ \nonumber
\sigma \left(f_0=\ar {\bf{q\bar{q}q\bar{q}}}\ra_{s-{\rm wave}} +\dots \right)       & \sim& \frac{\rm constant}{\bf s^5}   
\end{eqnarray}

If the experiment is carried out with insufficient energy to reveal the quark--gluon degrees of freedom, the two final state mesons effectively act as if pointlike so that its ``high--energy'' behavior differs from the QCD prediction: the left plot of figure~\ref{fig:Feynmanhadrons} reveals $n_f+L=2$ instead of 
$n_f+L\geq 4$ for composite hadrons. That entails that, if the cross section for $e^-e^+\to\phi f_0$ falls as $1/s$ or in any case slower than $1/s^3$  (with a small allowance for logarithmic corrections), the QCD degrees of freedom are not been extracted. Such $1/s$ behavior provides a baseline to judge whether the asymptotic regime is being seen.

The reaction is accessible to Belle-II. Data taken at 9 and 11 GeV would certainly satisfy the condition $s\gg (2 {\rm GeV})^2$ so the asymptotic power laws should be manifest.
The reaction cross sections at those two energies would be in the ratio, for each quark and gluon valence composition of the $f_0$,
$ \frac{\sigma(9{\rm GeV})}{\sigma(11{\rm GeV})} \simeq 3.4 \ (gg)\ ; \ 5 \ (q\bar{q})_{L=1}\ ; \ 
7.5\ (qq\bar{q}\bar{q})$, etc.

\begin{figure}[h]
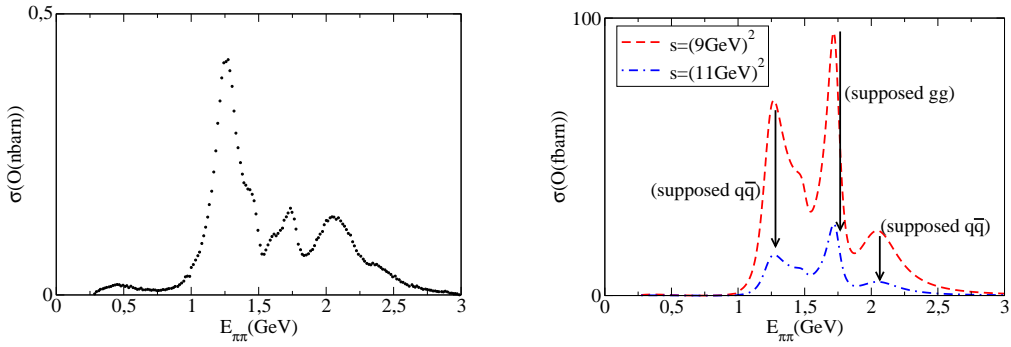

\begin{center}
\includegraphics*[width=0.4\columnwidth]{FIGS.DIR/Jpsitopipiphoton.eps}\hspace{1cm}
\includegraphics*[width=0.4\columnwidth]{FIGS.DIR/pipispectrum.eps}
\caption{ \label{fig:spectrumdistort} 
Left:  $\pi\pi$ spectrum at 3 GeV obtained~\cite{Bennett:2014fgt} from $J/\psi \gamma \pi\pi$ decays.
Right: $\pi\pi$ spectrum resulting from 
$e^-e^+\to \phi f_J$ at 9 and 11 GeV having assumed for the sake of exemplifying that $f_0(1710)$ 
would dominantly be the glueball (but whatever state or combination of states whose production rate 
would drop the least, if consistent with Eq.~(\protect\ref{glueballscaling}), would fit this assignment).
The normalization of the spectrum is fixed by actual Belle and Babar measurements of the reaction
$e^-e^+\to \phi \pi\pi$ at the $f_0(980)$ mass~\cite{Shen:2009mr}.
}
\end{center}
\end{figure}

Establishing that ratio with not too large an uncertainty, $3.4\pm 0.6$ (or the equivalent at another judicious choice of energy) would be the telltale of gluonium production. This could manifest itself as one of the narrow $f_0$ peaks in the spectrum maintaining a higher production rate than the others, that would fall faster (as illustrated in figure~\ref{fig:spectrumdistort} taken from our publication~\cite{Brodsky:2018snc}), or as a broad part of the spectrum behaving with such power--law.
 
The cross section at hard energies is small, $\sigma(9{\rm GeV})\sim70$ fbarn, but the extraordinary luminosity of Belle-II could yield 70000 $\phi$--recoiling $f_0(1710)$s with 1 ab$^{-1}$ of integrated luminosity (5 ab$^{-1}$ per year are planned with the accelerator working at design luminosity). 
About 20000 events are also obtainable at 11 GeV.

If a state appears to behave as a glueball according to the counting rules, its isoscalar gluonium nature can be further ascertained by verifying that no charged p-wave state with twist $\tau = 2$ appears at the same mass examining channels such as $e^+ e^- \to \rho^\pm  a^\mp$.

It is clear that Belle-II can impact hadron spectroscopy by identifying exotic mesons, including glueballs and tetraquarks~\cite{Drutskoy:2012gt,Kou:2018nap}.
By collecting significant off-resonance data at 9 and 11 GeV for example, 
it can  help with a longstanding puzzle, the identification of the glueball and generically classifying the $f_0$ mesons.
Moreover, any scalar meson $f_0$ which has an $O(1)$ mixing overlap with a glueball will have  $\sigma(e^+ e^- \to \phi f_0)$ scaling as $1/s^3$; thus, Belle can \emph{experimentally prove} the existence of a glueball even if it strongly mixed among several states, by just identifying a fraction of the spectrum with that specific scaling.

We have extended these arguments to other high momentum transfer exclusive and semi-inclusive reactions, for example $pp$ with both protons scattered elastically (e.g,  to roman pots set at fixed angles along the beam pipe) as in $pp\to pp \phi f_0$. 
If the mesons are deposited in the barrel with large transverse momentum (2-5 GeV for each meson) to suppress Regge exchanges, with all angular intervals fixed, we may apply the counting rules.
The two protons provide $n_i=6$ and an equal contribution to $n_f$.
The counting rules then predict  $d \sigma/dt \sim s^{2-n} = s^{-14}$   for $f_0\sim gg$,    
$s^{-15}$ for  $|q\bar{q}\ra$, and $s^{-16}$ for $|qq\bar{q}\bar{q}\ra$. 
The chances of distinguishing these steep and not so different fall-offs are small.

In consequence, doubly diffractive peripheral two-photon measurements with large $p_t$ (of several GeV for each of the two mesons in the barrel to suppress pomeron and reggeon exchanges in favor of photons) and a double gap to the diffracted protons looks more promising. 

Without the elastic suppression for the protons,  the prediction is identical to the one in $e^-e^+$ and $\gamma \gamma$ annihilation, Eq.~(\ref{glueballscaling}), since the effective reaction is 
$\gamma\gamma \to \phi f_0$:  $n_i=2$ instead of 6, and $n_f$ is due to the mesons only.
 
The power--law suppression of $d\sigma/dt$ is much less steep and more easily accessible than in the elastic case (though at fixed energy there are suppressing factors, the electric charge, the diffractive requirement on the protons, and the large $p_t$ requirement on the mesons).

Given the extensive body of work dedicated to gluonium, it comes as no surprise that there were earlier model computations of closely related processes, such as~\cite{Kivel:2017tgt} for the tensor glueball (a candidate to be on the pomeron Regge trajectory~\cite{Szczepaniak:1995cw});
the earlier~\cite{Wakely:1991ej} focused on $\gamma\gamma\to \eta[gg]$; 
and~\cite{Brodsky:2003hv} for $e^-e^+\to J/\psi f_0(gg)$, to name but a few. All these works seem to
have concentrated on specific width and production cross section calculations and have not highlighted the asymptotic QCD counting rules that provide model--independent access to the eventual existence and identification of gluonium.

Some other topics such as the wavefunction evolution from the hard experimental scale down to 3 GeV, the mixing of different configurations, etc. have been addressed in the companion publication~\cite{Brodsky:2018snc}.


\end{document}